\documentclass[
reprint,
superscriptaddress,
 amsmath,amssymb,
 aps,
 prl,
]{revtex4-1}

\usepackage{graphicx}
\usepackage{dcolumn}
\usepackage{bm}

\usepackage{soul}
\usepackage{xcolor}
\usepackage[colorlinks,
linkcolor=blue,
anchorcolor=blue,
citecolor=blue
]{hyperref}

\begin{document}

\preprint{APS/123-QED}

\title{Emergent spin-charge-orbital order in superconductor La$_3$Ni$_2$O$_7$}

\author{Binhua Zhang}
\affiliation{Key Laboratory of Computational Physical Sciences (Ministry of Education), Institute of Computational Physical Sciences, State Key Laboratory of Surface Physics, and Department of Physics, Fudan University, Shanghai 200433, China.}
\affiliation{Shanghai Qi Zhi Institute, Shanghai 200030, China.}

\author{Changsong Xu}
\email{csxu@fudan.edu.cn}
\affiliation{Key Laboratory of Computational Physical Sciences (Ministry of Education), Institute of Computational Physical Sciences, State Key Laboratory of Surface Physics, and Department of Physics, Fudan University, Shanghai 200433, China.}
\affiliation{Shanghai Qi Zhi Institute, Shanghai 200030, China.}

\author{Hongjun Xiang}
\email{hxiang@fudan.edu.cn}
\affiliation{Key Laboratory of Computational Physical Sciences (Ministry of Education), Institute of Computational Physical Sciences, State Key Laboratory of Surface Physics, and Department of Physics, Fudan University, Shanghai 200433, China.}
\affiliation{Shanghai Qi Zhi Institute, Shanghai 200030, China.}

\begin{abstract}
The bilayer nickelate La$_3$Ni$_2$O$_7$ (LNO) exhibits a remarkably high-temperature superconductivity of approximately 80 K under pressure, sparking considerable attentions. However, the nature of the spin, charge, and orbital order in LNO remains unknown, hindering the exploration of the mechanism of superconductivity. Here, supported by symmetry analysis and density functional theory calculations, we unravel
a double strip ground state of LNO with alternating magnetic moments arrangements. Interestingly, a charge density wave emerges under this determined magnetic order. Such CDW phase exhibits a Pmnm space group with breathing deformation of the NiO$_6$ octahedra, where the stretching and shrinking of the octahedra correspond to the formation of \emph{d}$_{z^2}^1$\emph{d}$_{x^2-y^2}^1$ and \emph{d}$_{z^2}^1$ orbital order. Moreover, we build a first-principles-based Hamiltonian for LNO, which provides deeper insight into its peculiar magnetic order. Our work thus reveals a systematic spin-charge-orbital picture for LNO, which can be extended to other nickelate-based superconductors, paving the way for determining the mechanism of superconductivity.
\end{abstract}

\maketitle

The quest to identify novel high-temperature superconductors  has significant implications for understanding the mechanisms of superconductivity and potentially enabling new applications. Recently, the Ruddlesden-Popper bilayer perovskite nickelate La$_3$Ni$_2$O$_7$ (LNO) was reported to feature a high superconducting transition temperature of approximately 80 K under pressure\cite{sun2023signatures}. This finding deviates from the copper-based paradigm, presenting a new promising platform to explore unconventional high-temperature superconductivity \cite{zhang2024high,zhu2023superconductivity,hou2023emergence,zhang2023electronic,sakakibara2024possible,lu2024interlayer,qu2024bilayer}.
  
\begin{figure*}[tbp]
  \centering
  \includegraphics[width=16cm]{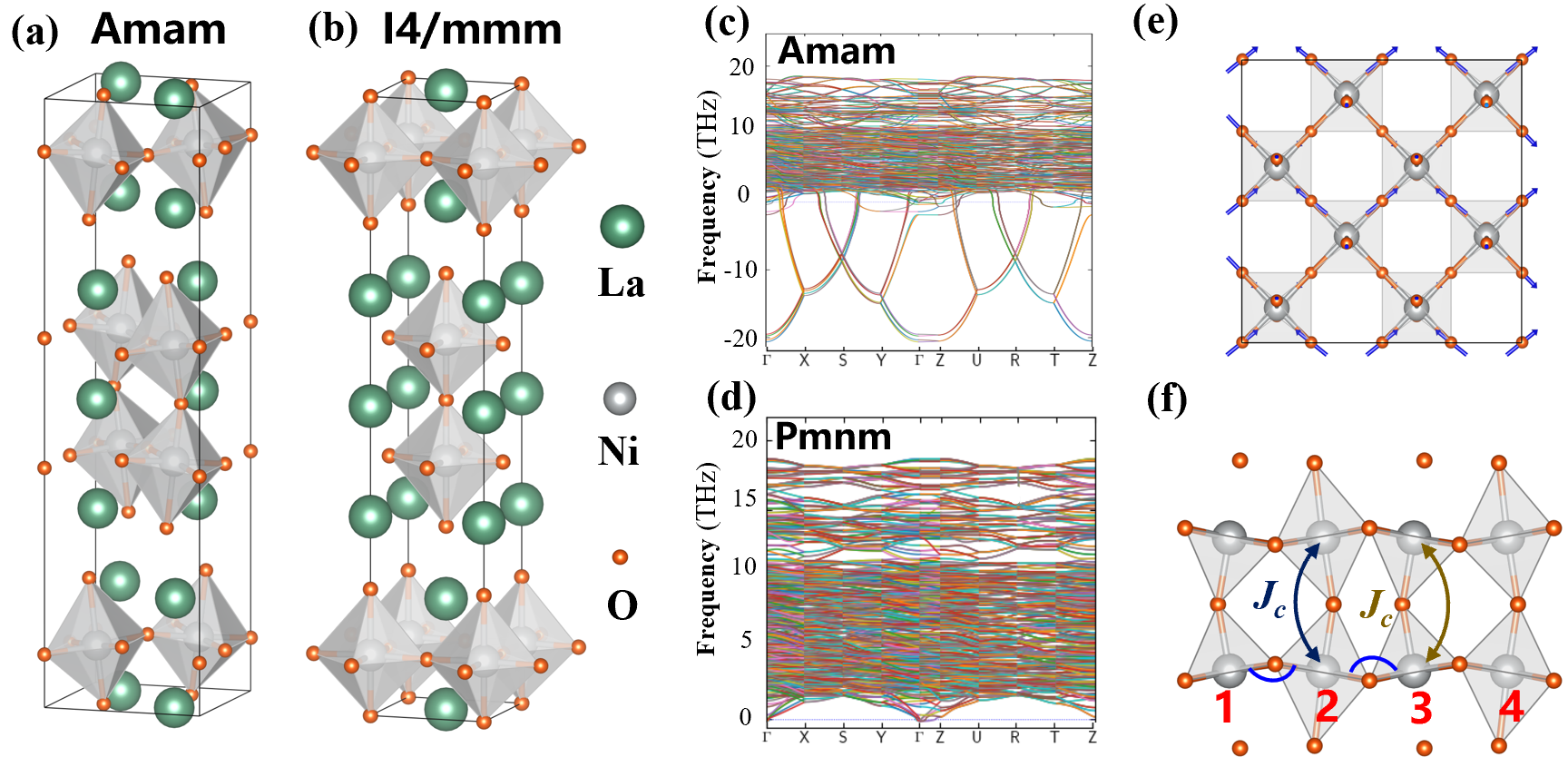}%
  \caption{Crystal structure and phonon spectrum. (a) Amam phase. (b) I4/mmm phase. The phonon dispersion of (c) Amam and (d) Pmnm phase of LNO with Eb1 magnetic order.(e) The breathing vibration mode at $\Gamma$ (0, 0, 0) point in (c). The blue arrows show the atomic displacements in this mode. (f) The Pmnm phase. To simplify the sketch only NiO$_6$ octahedra are displayed. The exchange paths are denoted by the solid double arrows.}
  \label{fig:structure}
\end{figure*}

LNO exhibits intricate quantum orders that are presumably intertwined with its superconductivity. At ambient pressure, LNO has been reported to present an Amam phase (No. 63) \cite{sun2023signatures,ling2000neutron,liu2023evidence}, which crystallizes into corner-connected NiO$_6$ octahedra with tilted distortion, as depicted in Fig. \ref{fig:structure}(a). Upon applying pressure, the superconductivity emerges, accompanied by the Ni-O-Ni bond angle changing from 168° to 180°, forming the higher-symmetry Fmmm or I4/mmm phase \cite{wang2024structure} [Fig. \ref{fig:structure}(b)]. There have been several investigations into the mechanism of superconductivity in LNO. For example, from an orbital perspective, the Ni-\emph{d}$_{z^2}$ orbital has been reported to be mainly relevant to the emergent superconductivity, as supported by pressurized electronic structure and spin susceptibility calculations \cite{sun2023signatures,zhang2024structural,yang2024orbital,luo2023bilayer,li2024electronic,christiansson2023correlated}. Moreover, the spin density wave was evidenced by recent nuclear magnetic resonance, resonant inelastic X-ray scattering, and muon spin relaxation measurements \cite{chen2024evidence,dan2024spin,chen2024electronic,khasanov2024pressure}, suggesting a competition between the spin density wave and superconductivity. However, despite firm experimental and theoretical explorations, the nature of the spin, charge, and orbital order in LNO remains unknown, leaving the mechanism of superconductivity elusive.
  
  In this work, we employ symmetry analysis to generate all non-equivalent antiferromagnetic (AFM) configurations in the 2 $\times$ 2 $\times$ 1 supercells of LNO at ambient pressure.  By combining density functional theory (DFT) calculations, we determine that the ground state is a double stripe configuration, characterized by the spins of the Ni atoms aligning ferromagnetically in the \emph{a}-direction and alternating antiferromagnetically in the \emph{b}-direction. Such configuration also exhibits alternating magnetic moment arrangements and is suppressed with increasing pressure. Particularly, we find a charge density wave (CDW) transition related to the breathing deformation of the NiO$_6$ octahedra under the peculiar magnetic ground state. We further demonstrate that the newly discovered CDW phase shares the Pmnm (No. 58) space group, characterized by alternate stretching and shrinking of the octahedra, thereby forming the \emph{d}$_{z^2}^1$\emph{d}$_{x^2-y^2}^1$ and \emph{d}$_{z^2}^1$ orbital order, respectively. Moreover, based on spin Hamiltonian, we reveal the ferromagnetic (FM) first nearest neighbor interaction and AFM second nearest neighbor interaction, which may result from the inteplay between double exchange and superexchange. The interlayer coupling exhibits distinct AFM coupling strengths, attributed to the alternating orbital arrangement. Our work thus sheds light on the intertwined spin-charge-orbital interactions in LNO, potentially paving the way for exploring the mechanism of superconductivity.
  
\begin{figure*}[tb]
  \centering
  \includegraphics[width=16cm]{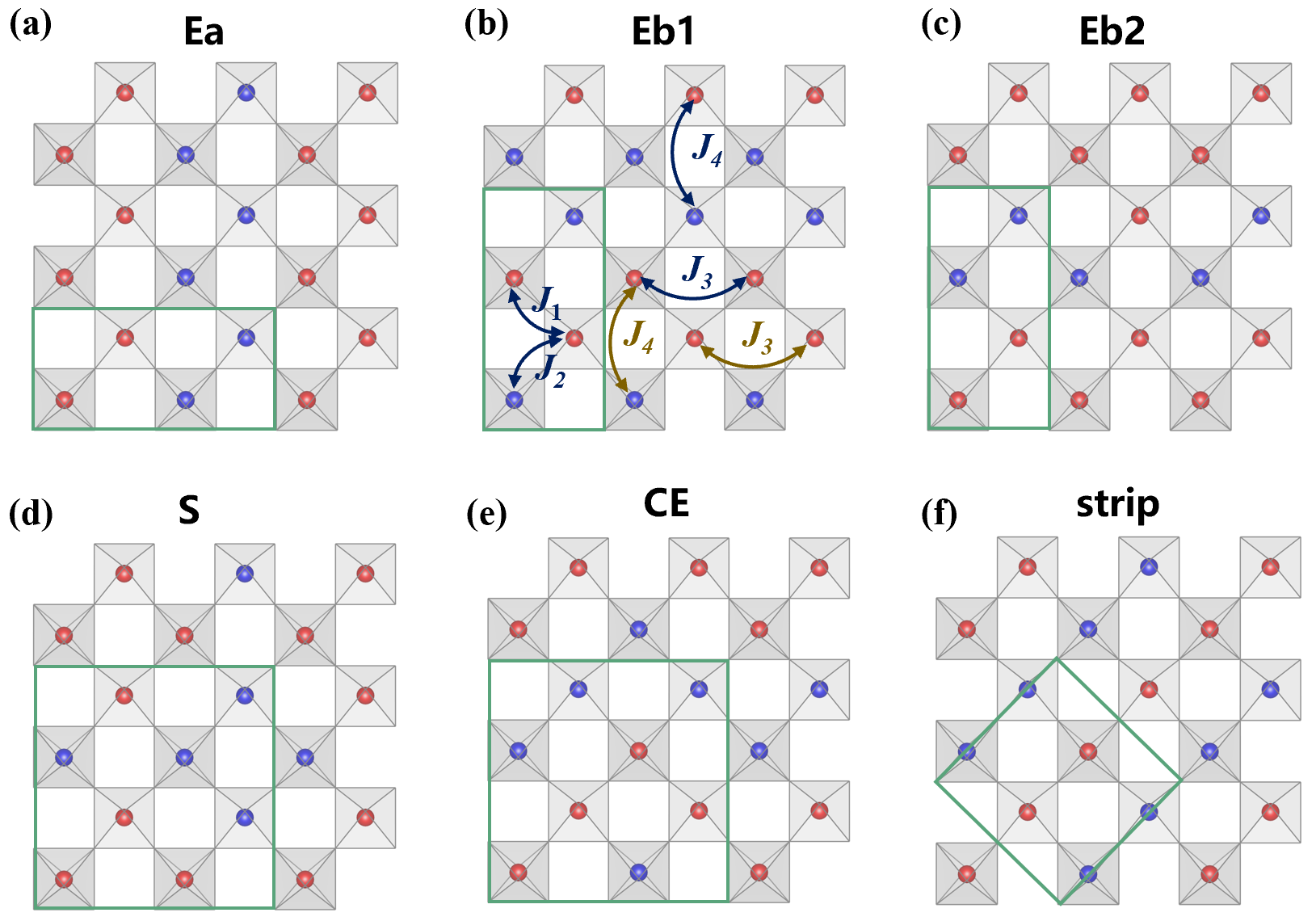}%
  \caption{The possible magnetic configurations of LNO.  To simplify the AFM sketch, only one layer of nickel cations is displayed. Here, the red and blue balls represent spin up and spin down, respectively. Green solid lines mark the unit cell of magnetic order. The exchange paths are denoted by the solid double arrows.}
  \label{fig:magnetic}
\end{figure*}

\textcolor{blue}{Spin density wave.} To elucidate the possible magnetic order of LNO at ambient pressure, we employ a symmetry analysis method to generate all non-equivalent AFM configurations within the 2 $\times$ 2 $\times$ 1 supercells of the Amam phase. Such a method results in a total of 331 distinct magnetic configurations, including the double-strip pattern [Figs. \ref{fig:magnetic}(a-c)], the square-convergent pattern [Fig.\ref{fig:magnetic}(d)], the zigzag FM chain pattern [Fig.\ref{fig:magnetic}(e)], the stripe pattern [Fig.\ref{fig:magnetic}(f)], typical A-, C- and G-type pattern, etc. Note that the three double-stripe patterns depicted in Figs. \ref{fig:magnetic}(a-c)], with Ni atoms aligning ferromagnetically in different directions and sites, are not equivalent in symmetry.

\begin{figure*}[tbp]
  \centering
  \includegraphics[width=16cm]{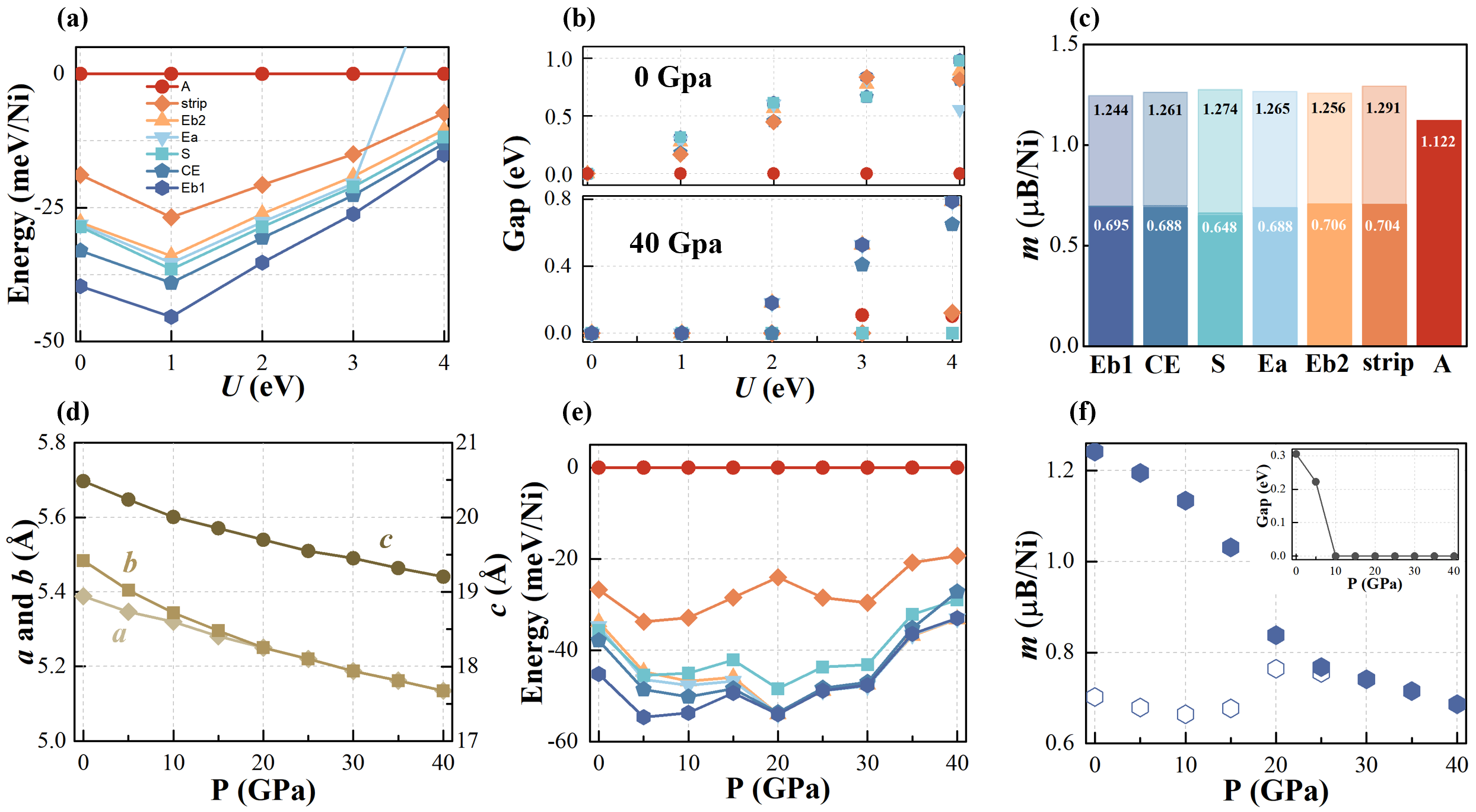}%
  \caption{Magnetic order under DFT + \emph{U} calculations and the effect of pressure. (a) The dependence of energy on effective on-site Hubbard interactions \emph{U} for different magnetic orders relative to the A-type structure. (b) The band gap of LNO at 0 Gap and 40 Gpa under different \emph{U} values. (c) The distribution of magnetic moments for different configurations at \emph{U} = 1 eV. (d) The evolution of lattice constants under pressure. (e) The relative total energy of different magnetic orders under pressure. (f) The change in magnetic moments of the Eb1 state with pressure. The inset shows the corresponding band gap evolution.}
  \label{fig:U}
\end{figure*}

 To further determine the magnetic ground state, we employed the DFT+U method (U$_{eff}$ =1 eV) for all aforementioned AFM magnetic configurations, using the experimental lattice constants (\emph{a} = 5.4018 \AA, \emph{b} = 5.4557 \AA, \emph{c} = 20.537 \AA). As shown in Fig. \ref{fig:U}(a), the double stripe configuration, characterized by the spins of the Ni atoms aligning ferromagnetically in the \emph{a}-direction and alternating antiferromagnetically in the \emph{b}-direction [Eb1 state in Fig.\ref{fig:magnetic}(b)],is more energetically favorable, distinguishing it from the Eb2 and Ea states [Fig.\ref{fig:magnetic}(a,c)]. The CE configuration is slightly higher than the Eb1 state, with an energy difference of 10 meV/Ni. Note that all of the Eb1, Eb2, Ea and CE configurations align with the resonant inelastic X-ray scattering and nuclear magnetic resonance measurements \cite{chen2024electronic,dan2024spin}.

The peculiar magnetic order in LNO also induces a rearrangement of magnetic moments, which is essential for understanding its spin density wave. As shown in Fig. \ref{fig:U}(c), all configurations presented in Fig. \ref{fig:magnetic} exhibit a similar alignment of magnetic moments, with moments of $\sim$ 0.69 $\mu$B and $\sim$ 1.24 $\mu$B alternating in a 1 $\times$ 1 $\times$ 1 supercell. However, the typical A-type AFM configuration, as well as G- and C- types cases, feature only one mixed magnetic moment. This is understandable since the latter configurations involve only one non-equivalent Ni site based on symmetry analysis, whereas the former involves at least two Ni sites.

We then look at the effects of pressure since superconductivity emerges upon applying pressure. As shown in Fig. \ref{fig:U}(d), the low-symmetry phase with tilted NiO$_6$ octahedra gradually transforms into the undistorted phase as pressure increases, reaching the high-symmetry I4/mmm phase at 20 GPa, consistent with experimental observations. Correspondingly, LNO has an Eb1 ground state at low pressure. As pressure increase, the Eb1 state becomes equivalent to the Eb2 and Ea states in the I4/mmm phase at 20 GPa, as clearly captured in Fig. \ref{fig:U}(e), aligning with symmetry. In addition, the band gap in the Eb1 state disappears at 10 GPa, causing the system to transform from a weakly insulating to a metallic phase. Noteworthily, the magnetic moments in the Eb1 state also gradually decrease with increasing pressure. Especially the alternating magnetic moments are suppressed and merge at high pressure. It is thus reasonable to speculate that superconductivity competes with the spin density wave and can be be achieved by suppressing spin density through the application of pressure.

Moreover, we tested the effects of effective on-site Hubbard interactions (U). As shown in Fig. \ref{fig:U}(a), the Eb1 state is always energetically favorable within a range of U values from 0 to 4 eV. However, large U values will cause the system to remain an insulator even in the high-pressure I4/mmm phase [Fig. \ref{fig:U}(b)]. Considering the superconductivity under high pressure, small U values ( U\textless 2 eV ) may be more suitable for the present LNO system. Therefore, the U value of 1 eV we adopted in most calculations meets these requirements.

\textcolor{blue}{Charge density wave.} We now investigate the hidden structure instability using the aforementioned Eb1 magnetic ground state. Phonon dispersion of the Amam phase at zero pressure is displayed in Fig. \ref{fig:structure}(c). The
 imaginary phonon mode at the $\Gamma$ point 
 corresponds to a breathing motion
of the NiO$_6$ octahedra, characterized by alternate stretching and shrinking of the octahedra, as shown in Figs. \ref{fig:structure}(e). Superposing this mode onto the pristine Amam phase leads to a new 1$\times$1$\times$1 structural reconstruction, possessing the space group of Pmnm (No. 58). The optimized Pmnm phase is dynamically stable, where the breathing deformation reduces the total energy. The imaginary frequencies disappear in its phonon spectra, as shown in Fig. \ref{fig:structure}(d). Note that the new Pmnm phase has two inequivalent Ni sites (i.e., Ni1 and Ni2), as illustrated in Fig. \ref{fig:structure}(f), which correspond to aforementioned alternating magnetic moments. The Eb1 magnetic ground state holds 1$\times$2$\times$1 supercells of Pmnm phase, resulting in \emph{q}$_{cdw}$=2\emph{q}$_{sdw}$ (here \emph{q} denotes propagation vectors), consistent with reported La$_4$Ni$_3$O$_{10}$ \cite{zhang2020intertwined}.

\begin{figure*}[tbp]
  \centering
  \includegraphics[width=16cm]{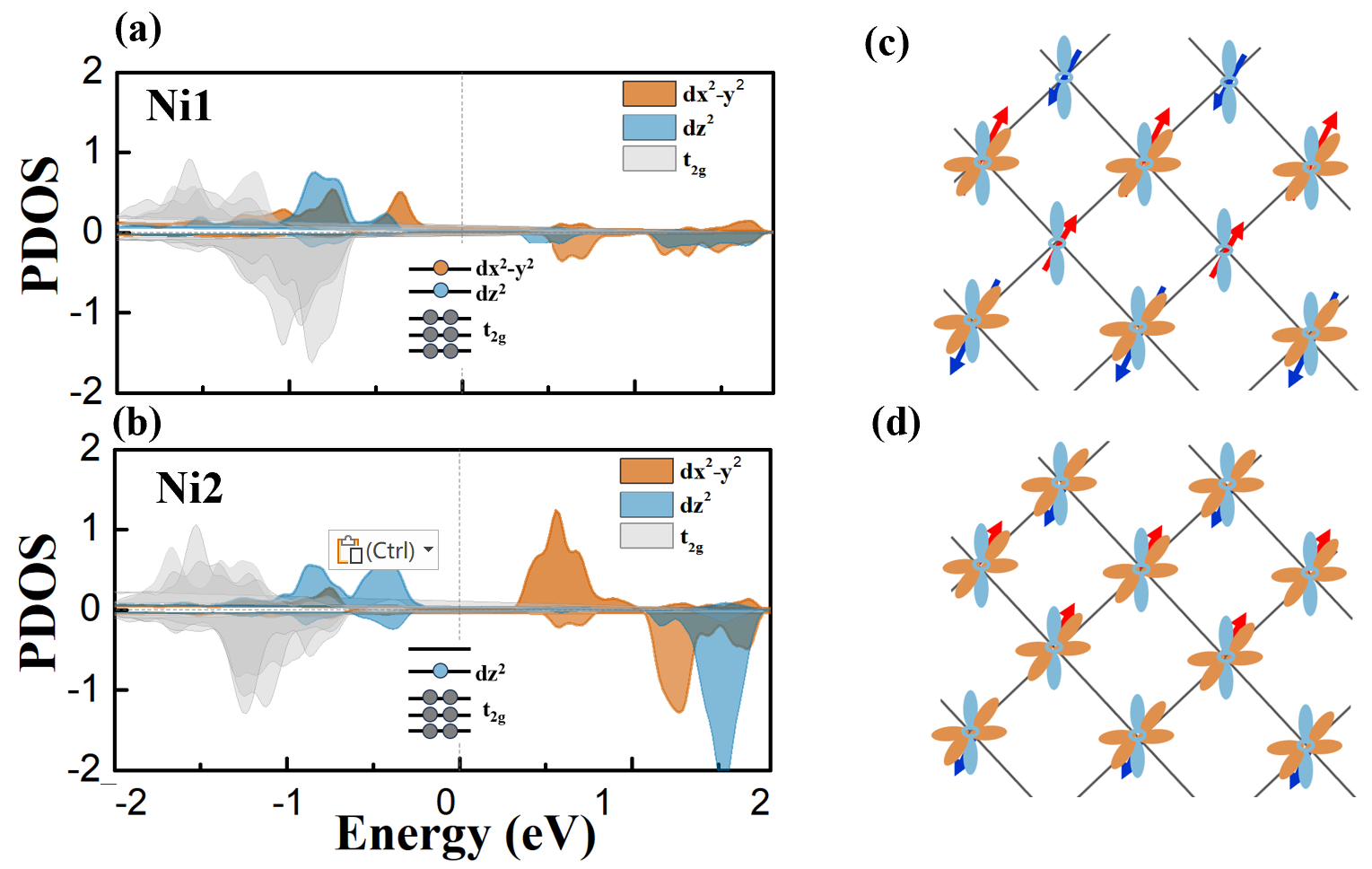}%
  \caption{Orbital order. The projected densities of states (PDOS) for (a) Ni1 and (b) Ni2 in the Eb1 state of Pmnm phase. Here we only display the spin-up case. Insets show the schematic of electron configuration of $3d^8$ and $3d^7$. (c) Alternating orbital order at Pmnm phase. This orbital arrangement is plotted based on the calculated charge density distribution for the energy window from -0.6 eV to 0 eV in (a) and (b). (d) The average mixed orbital in the pressurized I4/mmm phase.}
  \label{fig:orbital}
\end{figure*}

\textcolor{blue}{Orbital order.}
The intertwined spin and charge density waves naturally accompany the orbital order. For the Ni1 atom, with stretching NiO$_6$ octahedra, the splitting between the \emph{d}$_{z^2}$ and \emph{d}$_{x^2-y^2}$ orbitals is weak. As a result, Ni1 has a valence state of +2 with a 3\emph{d}$^8$ (\emph{t}$_{2g}^6$\emph{d}$_{z^2}^1$\emph{d}$_{x^2-y^2}^1$) electron configuration, as plotted in Fig. \ref{fig:orbital}(a). In contrast, the Ni2 atom, characterized by the shrinking of the NiO$_6$ octahedra, manifests a large splitting of the \emph{e}$_g$ orbitals, so only the \emph{d}$_{z^2}$ is occupied due to the Jahn-Teller effect, forming Ni$^{3+}$ with a 3\emph{d}$^7$(\emph{t}$_{2g}^6$\emph{d}$_{z^2}^1$) electron configuration, in line with the PDOS in \ref{fig:orbital}(b). The alternating Ni$^{2+}$ and Ni$^{3+}$ atoms, accompanied by alternating magnetic moments ($\sim$ 1.24 $\mu$B and $\sim$ 0.69 $\mu$B, respectively), form the regular orbital order in LNO, as depicted in Fig. \ref{fig:orbital}(c) (see calculated electron density distribution in SM \cite{SM}). Note that the alternating orbital order of \emph{d}$_{z^2}^1$\emph{d}$_{x^2-y^2}^1$ and \emph{d}$_{z^2}^1$ is gradually suppressed under pressure, achieving one average mixed orbital in the high-symmetry phase without breathing distortion [Fig. \ref{fig:orbital}(d)]. 

\textcolor{blue}{Magnetic interaction.} We now work on understanding the peculiar Eb1 spin arrangement in the Pmnm phase of LNO. We build a first-principle-based spin
 Hamiltonian for LNO, reads
\begin{equation}
\begin{aligned}
\mathcal{H}= &\sum_{\langle i,j \rangle_n}J_nS_iS_j + \sum_{\langle i,j \rangle_\perp}J_\perp S_iS_j 
\end{aligned}
\label{eq1}
\end{equation}
where $\langle i,j \rangle_n$ denotes pairs of nth
 nearest neighbors (NN) within each layer, while the $\perp$
 symbol refers to interlayer couplings; \emph{J} quantifies magnetic exchange interactions. With input energies of all possible AFM structures from DFT calculations, the coefficients of these exchange interaction can be fitted using the machine learning method for constructing the Hamiltonian. Note that there are a total of 331 nonequivalent configurations; here, we select only the configurations with similar alternating magnetic arrangements where the change in magnetic moment is less than 0.1 $\mu$B ($\sim$ 80 structures).

 \begin{table}[h]
 \caption{Magnetic parameters of Eq. \ref{eq1} fitted from different \emph{U} values, in unit of meV. The positive and negative signs indicate the AFM and FM interactions, respectively. the $\perp$ symbol refers to interlayer couplings. The exchange paths are shown in Fig. \ref{fig:structure}(f) and Fig. \ref{fig:magnetic}(b). Note that the split exchange coupling strength are listed in parentheses.}
\begin{ruledtabular}
\begin{tabular}{c c c c}
\textrm{LNO}&
\textrm{Length (\AA)}&
\textrm{\emph{U} = 1 eV}&
\textrm{\emph{U} = 2 eV}\\
\colrule

$J_1$ &	 3.81   & 	-1.1 &	-4.4\\
$J_2$ & 3.87 &		6.6 &	1.9\\
$J_\perp$	& 3.93 &	124.1 (43.3) &	122.6 (33.7)\\
$J_3$ & 5.40 &		2.9 (-6.2) &	-0.8 (-3.3)\\
$J_4$ &	5.46 &	1.6 (2.7) &	-1.9 (3.8)\\
\end{tabular}
\end{ruledtabular}
\label{J}
\end{table}
 
 As shown in Table \ref{J}, the 1NN exchange interaction favors FM with \emph{J}$_1$ = -1.1 meV, which may stem from the dominant double exchange between \emph{d}$_{z^2}^1$\emph{d}$_{x^2-y^2}^1$ and \emph{d}$_{z^2}^1$ orbitals. The 2NN interaction overcomes the double exchange due to enhanced Ni-O-Ni superexchange, yielding AFM with \emph{J}$_2$ = 6.6 meV. Based on such \emph{J}$_1$ and \emph{J}$_2$ values, it is clear to see that the Eb1 state is more energetically favorable than the Eb2 state, with an energy gain of 2\emph{J}$_1$-2\emph{J}$_1$. 
 On the other hand, \emph{J}$_\perp$ favors strong AFM with alternating values of 124 meV and 43 meV, whose average aligns with experimental estimations \cite{chen2024electronic}. This is understandable since the alternating orbital arrangement splits the interlayer coupling into two paths, namely the \emph{d}$_{z^2}^1$\emph{d}$_{x^2-y^2}^1$-\emph{d}$_{z^2}^1$\emph{d}$_{x^2-y^2}^1$ path and \emph{d}$_{z^2}^1$-\emph{d}$_{z^2}^1$ path, resulting in distinct coupling strengths. Likewise, the in-plane 3NN and 4NN interactions also exhibit split values under these two paths, which may cooperate with 1NN and 2NN to further stabilize the Eb1 state. Additionally, Monte Carlo simulations with these obtained Hamiltonians indeed yield the Eb1 ground state, revealing a good agreement with DFT calculations.

 In summary, combing symmetry analysis and DFT calculations, we reveal the Eb1 ground state of LNO with alternating magnetic moments arrangements. We point out that this peculiar magnetic order induces a structural instability, lead to a new CDW phase. Such CDW phase exhibits a Pmnm space group with breathing deformation of the NiO$_6$ octahedra. The stretching and shrinking of the octahedra correspond to the formation of the 3\emph{d}$^7$ and 3\emph{d}$^8$ electron configuration under the Jahn-Teller effect. Moreover, we construct the spin Hamiltonian for LNO, which may provide deeper insight into its magnetic properties. Experimental verification of the emergent CDW state and intertwined spin-charge-orbital picture revealed in this work is called for.

\begin{acknowledgments}
  We acknowledge financial support from the National Key R\&D Program of China (No. 2022YFA1402901), the support from NSFC (grants No. 11991061, 12188101, 12174060, and 12274082), and the Guangdong Major Project of the Basic and Applied Basic Research (Future functional materials under extreme conditions--2021B0301030005). C. X. also acknowledge the support from Shanghai Science and Technology Committee (grant No. 23ZR1406600). B. Z. also acknowledge the support from China Postdoctoral Science Foundation (grants No. 2024T170152, 2022M720816).
\end{acknowledgments}

\newpage{\pagestyle{empty}\cleardoublepage}


\begin{thebibliography}{100}

	\bibitem{sun2023signatures} H.~Sun, M.~Huo, X.~Hu, J.~Li, Z.~Liu, Y.~Han, L.~Tang, Z.~Mao, P.~Yang, B.~Wang \emph{et~al.}, ``Signatures of superconductivity near 80 k in a nickelate under high pressure,'' \emph{Nature}, vol. 621, no. 7979, pp. 493--498, 2023.

	\bibitem{zhang2024high} Y.~Zhang, D.~Su, Y.~Huang, Z.~Shan, H.~Sun, M.~Huo, K.~Ye, J.~Zhang, Z.~Yang, Y.~Xu \emph{et~al.}, ``High-temperature superconductivity with zero resistance and strange-metal behaviour in la3ni2o7- $\delta$,'' \emph{Nature Physics}, pp. 1--5, 2024.

	\bibitem{zhu2023superconductivity} Y.~Zhu, E.~Zhang, B.~Pan, X.~Chen, D.~Peng, L.~Chen, H.~Ren, F.~Liu, N.~Li, Z.~Xing \emph{et~al.}, ``Superconductivity in trilayer nickelate la $ \_4 $ ni $ \_3 $ o $ \_ $\{$10$\}$ $ single crystals,'' \emph{Nature}, vol. 631, p. 531–536, 2024.

	\bibitem{hou2023emergence} J.~Hou, P.-T. Yang, Z.-Y. Liu, J.-Y. Li, P.-F. Shan, L.~Ma, G.~Wang, N.-N. Wang, H.-Z. Guo, J.-P. Sun \emph{et~al.}, ``Emergence of high-temperature superconducting phase in pressurized la3ni2o7 crystals,'' \emph{Chinese Physics Letters}, vol.~40, no.~11, p. 117302, 2023.

	\bibitem{zhang2023electronic} Y.~Zhang, L.-F. Lin, A.~Moreo, and E.~Dagotto, ``Electronic structure, dimer physics, orbital-selective behavior, and magnetic tendencies in the bilayer nickelate superconductor la 3 ni 2 o 7 under pressure,'' \emph{Physical Review B}, vol. 108, no.~18, p. L180510, 2023.

	\bibitem{sakakibara2024possible} H.~Sakakibara, N.~Kitamine, M.~Ochi, and K.~Kuroki, ``Possible high t c superconductivity in la 3 ni 2 o 7 under high pressure through manifestation of a nearly half-filled bilayer hubbard model,'' \emph{Physical Review Letters}, vol. 132, no.~10, p. 106002, 2024.

	\bibitem{lu2024interlayer} C.~Lu, Z.~Pan, F.~Yang, and C.~Wu, ``Interlayer-coupling-driven high-temperature superconductivity in la 3 ni 2 o 7 under pressure,'' \emph{Physical Review Letters}, vol. 132, no.~14, p. 146002, 2024.

	\bibitem{qu2024bilayer} X.-Z. Qu, D.-W. Qu, J.~Chen, C.~Wu, F.~Yang, W.~Li, and G.~Su, ``Bilayer t-j-j$\perp$ model and magnetically mediated pairing in the pressurized nickelate La$_3$Ni$_2$O$_7$,'' \emph{Physical Review Letters}, vol. 132, no.~3, p. 036502, 2024.

	\bibitem{ling2000neutron} C.~D. Ling, D.~N. Argyriou, G.~Wu, and J.~Neumeier, ``Neutron diffraction study of la3ni2o7: Structural relationships among n= 1, 2, and 3 phases lan+ 1nino3n+ 1,'' \emph{Journal of Solid State Chemistry}, vol. 152, no.~2, pp. 517--525, 2000.

	\bibitem{liu2023evidence} Z.~Liu, H.~Sun, M.~Huo, X.~Ma, Y.~Ji, E.~Yi, L.~Li, H.~Liu, J.~Yu, Z.~Zhang \emph{et~al.}, ``Evidence for charge and spin density waves in single crystals of la3ni2o7 and la3ni2o6,'' \emph{Science China Physics, Mechanics \& Astronomy}, vol.~66, no.~1, p. 217411, 2023.

	\bibitem{wang2024structure} L.~Wang, Y.~Li, S.-Y. Xie, F.~Liu, H.~Sun, C.~Huang, Y.~Gao, T.~Nakagawa, B.~Fu, B.~Dong \emph{et~al.}, ``Structure responsible for the superconducting state in la3ni2o7 at high-pressure and low-temperature conditions,'' \emph{Journal of the American Chemical Society}, vol. 146, no.~11, pp. 7506--7514, 2024.

	\bibitem{zhang2024structural} Y.~Zhang, L.-F. Lin, A.~Moreo, T.~A. Maier, and E.~Dagotto, ``Structural phase transition, s$\pm$-wave pairing, and magnetic stripe order in bilayered superconductor la3ni2o7 under pressure,'' \emph{Nature Communications}, vol.~15, no.~1, p. 2470, 2024.

	\bibitem{yang2024orbital} J.~Yang, H.~Sun, X.~Hu, Y.~Xie, T.~Miao, H.~Luo, H.~Chen, B.~Liang, W.~Zhu, G.~Qu \emph{et~al.}, ``Orbital-dependent electron correlation in double-layer nickelate la3ni2o7,'' \emph{Nature Communications}, vol.~15, no.~1, p. 4373, 2024.

	\bibitem{luo2023bilayer} Z.~Luo, X.~Hu, M.~Wang, W.~W{\'u}, and D.-X. Yao, ``Bilayer two-orbital model of l a 3 n i 2 o 7 under pressure,'' \emph{Physical review letters}, vol. 131, no.~12, p. 126001, 2023.

	\bibitem{li2024electronic} Y.~Li, X.~Du, Y.~Cao, C.~Pei, M.~Zhang, W.~Zhao, K.~Zhai, R.~Xu, Z.~Liu, Z.~Li \emph{et~al.}, ``Electronic correlation and pseudogap-like behavior of high-temperature superconductor la3ni2o7,'' \emph{Chinese Physics Letters}, 2024.

	\bibitem{christiansson2023correlated} V.~Christiansson, F.~Petocchi, and P.~Werner, ``Correlated electronic structure of la 3 ni 2 o 7 under pressure,'' \emph{Physical Review Letters}, vol. 131, no.~20, p. 206501, 2023.

	\bibitem{chen2024evidence} K.~Chen, X.~Liu, J.~Jiao, M.~Zou, C.~Jiang, X.~Li, Y.~Luo, Q.~Wu, N.~Zhang, Y.~Guo \emph{et~al.}, ``Evidence of spin density waves in la 3 ni 2 o 7-$\delta$,'' \emph{Physical Review Letters}, vol. 132, no.~25, p. 256503, 2024.

	\bibitem{dan2024spin} Z.~Dan, Y.~Zhou, M.~Huo, Y.~Wang, L.~Nie, M.~Wang, T.~Wu, and X.~Chen, ``Spin-density-wave transition in double-layer nickelate la3ni2o7,'' \emph{arXiv preprint arXiv:2402.03952}, 2024.

	\bibitem{chen2024electronic} X.~Chen, J.~Choi, Z.~Jiang, J.~Mei, K.~Jiang, J.~Li, S.~Agrestini, M.~Garcia-Fernandez, H.~Sun \emph{et~al.}, ``Electronic and magnetic excitations in la3ni2o7,'' 2024.

	\bibitem{khasanov2024pressure} R.~Khasanov, T.~J. Hicken, D.~J. Gawryluk, L.~P. Sorel, S.~B{\"o}tzel, F.~Lechermann, I.~M. Eremin, H.~Luetkens, and Z.~Guguchia, ``Pressure-induced split of the density wave transitions in La$_3$Ni$_2$O$_{7-\delta}$,'' \emph{arXiv preprint arXiv:2402.10485}, 2024.

	\bibitem{zhang2020intertwined} J.~Zhang, D.~Phelan, A.~Botana, Y.-S. Chen, H.~Zheng, M.~Krogstad, S.~G. Wang, Y.~Qiu, J.~Rodriguez-Rivera, R.~Osborn \emph{et~al.}, ``Intertwined density waves in a metallic nickelate,'' \emph{Nature communications}, vol.~11, no.~1, p. 6003, 2020.

	\bibitem{SM} See Supplemental Material at http://link for more information, which includes {Refs}. \cite{perdew1996generalized,kresse1996efficiency,kresse1996efficient2,blochl1994projector,grimme2010consistent,togo2015first,wang2021vaspkit}.

	\bibitem{yu2024density} X.~H. Yu, Y.~Meng, Y.~Yang, H.~Sun, S.~Zhang, J.~Luo, L.~Chen, X.~Ma, M.~Wang, F.~Hong \emph{et~al.}, ``Density-wave-like gap evolution in la3ni2o7 under high pressure revealed by ultrafast optical spectroscopy,'' 2024.

	\bibitem{liu2023s} Y.-B. Liu, J.-W. Mei, F.~Ye, W.-Q. Chen, and F.~Yang, ``s$\pm$-wave pairing and the destructive role of apical-oxygen deficiencies in la 3 ni 2 o 7 under pressure,'' \emph{Physical Review Letters}, vol. 131, no.~23, p. 236002, 2023.

	\bibitem{wang2021vaspkit} V.~Wang, N.~Xu, J.-C. Liu, G.~Tang, and W.-T. Geng, ``{VASPKIT}: A user-friendly interface facilitating high-throughput computing and analysis using {VASP} code,'' \emph{Comput. Phys. Commun.}, vol. 267, p. 108033, 2021.

	\bibitem{perdew1996generalized} J.~P. Perdew, K.~Burke, and M.~Ernzerhof, ``Generalized gradient approximation made simple,'' \emph{Phys. Rev. Lett.}, vol.~77, no.~18, p. 3865, 1996.

	\bibitem{kresse1996efficiency} G.~Kresse and J.~Furthm{\"u}ller, ``Efficiency of \emph{ab-initio} total energy calculations for metals and semiconductors using a plane-wave basis set,'' \emph{Comput. Mater. Sci.}, vol.~6, no.~1, pp. 15--50, 1996.

	\bibitem{kresse1996efficient2} G.~Kresse and J.~Furthm\"uller, ``Efficient iterative schemes for \emph{ab initio} total-energy calculations using a plane-wave basis set,'' \emph{Phys. Rev. B}, vol.~54, no.~16, p. 11169, 1996.

	\bibitem{blochl1994projector} P.~E. Bl{\"o}chl, ``Projector augmented-wave method,'' \emph{Phys. Rev. B}, vol.~50, no.~24, p. 17953, 1994.

	\bibitem{grimme2010consistent} S.~Grimme, J.~Antony, S.~Ehrlich, and H.~Krieg, ``A consistent and accurate ab initio parametrization of density functional dispersion correction {(DFT-D)} for the 94 elements {H-Pu},'' \emph{J. Chem. Phys.}, vol. 132, no.~15, p. 154104, 2010.

	\bibitem{togo2015first} A.~Togo and I.~Tanaka, ``First principles phonon calculations in materials science,'' \emph{Scr. Mater.}, vol. 108, pp. 1--5, 2015.


\end{thebibliography}
\end{document}